\documentclass[aps,prd,preprint,superscriptaddress,tightenlines,nofootinbib,showpacs]{revtex4}

\usepackage{graphicx}
\usepackage{dcolumn}
\usepackage{bm}

\def\Y{\ifmmode \Upsilon \else%
$\Upsilon$ %
\fi}
\def\chib{\ifmmode \chi_b \else%
$\chi_b$ %
\fi}
\def\chibp{\ifmmode \chi_b' \else%
$\chi_b$ %
\fi}
\def\Q#1#2#3#4{\ifmmode
 \,#1\,{^{#2}#3}_{#4}
\else%
$#1\,{^{#2}#3}_{#4}$ %
\fi}
\def\eonem#1#2{\ifmmode
\left| <#1|r|#2> \right|
\else%
$\left| <#1|r|#2> \right|$
\fi}

\def\ee{\ifmmode e^+e^- \else $e^+e^-$  \fi}
\def\mm{\ifmmode \mu^+\mu^- \else $\mu^+\mu^-$  \fi}
\def\LL{\ifmmode l^+l^- \else $l^+l^-$  \fi}

\begin{document}

\preprint{CLNS 04/1886}       
\preprint{CLEO 04-10}         

\title{Photon Transitions in $\psi(2S)$ Decays to $\chi_{cJ}(1P)$ and $\eta_c(1S)$}


\author{S.~B.~Athar}
\author{P.~Avery}
\author{L.~Breva-Newell}
\author{R.~Patel}
\author{V.~Potlia}
\author{H.~Stoeck}
\author{J.~Yelton}
\affiliation{University of Florida, Gainesville, Florida 32611}
\author{P.~Rubin}
\affiliation{George Mason University, Fairfax, Virginia 22030}
\author{B.~I.~Eisenstein}
\author{G.~D.~Gollin}
\author{I.~Karliner}
\author{D.~Kim}
\author{N.~Lowrey}
\author{P.~Naik}
\author{C.~Sedlack}
\author{M.~Selen}
\author{J.~J.~Thaler}
\author{J.~Williams}
\author{J.~Wiss}
\affiliation{University of Illinois, Urbana-Champaign, Illinois 61801}
% ...
\collaboration{CLEO Collaboration} 
\noaffiliation



\date{\today}

\begin{abstract} 
We have studied 
the inclusive photon spectrum in $\psi(2S)$ decays 
using the CLEO III detector.
We present the most precise measurements of 
electric dipole (E1) photon transition rates for
$\psi(2S)\to\gamma\chi_{cJ}(1P)$ ($J=0,1,2$).
We also confirm the hindered magnetic dipole (M1) transition, 
$\psi(2S)\to\gamma\eta_c(1S)$.
However, the direct M1 transition $\psi(2S)\to\gamma\eta_c(2S)$
observed by the Crystal Ball as a narrow peak at a photon energy of
$91$ MeV is not found in our data. 
\end{abstract}

\maketitle

Observation of the triplet of $\chi_{cJ}(1P)$  states ($J=0,1,2$) via radiative 
E1 transitions from $\psi(2S)$ \cite{chisPreCB} confirmed
the interpretation of the $J/\psi(1S)$ and $\psi(2S)$ states as
non-relativistic bound states of a heavy quark-antiquark system 
and solidified the quark model of hadrons.
The best exploration of the inclusive photon spectrum in $\psi(2S)$ decays 
was performed by the Crystal Ball experiment two decades ago \cite{CB86}.
The Crystal Ball also claimed observation of two singlet states, 
$\eta_c(1S)$ and $\eta_c(2S)$ via rare M1 photon transitions
\cite{CB82,CB86}.
Although many other experiments \cite{PDG} have confirmed 
the $\eta_c(1S)$ state, 
the Crystal Ball $\eta_c(2S)$ candidate 
remains the sole evidence for this latter state
at a mass of 3592 MeV corresponding to a photon energy of $91$ MeV. 
Searches for it in $\bar pp$ formation 
have been unsuccessful \cite{etacpFail}. 
Moreover, the validity of the Crystal Ball $\eta_c(2S)$ candidate has been put 
in serious doubt by the Belle experiment, which found
evidence for the $\eta_c(2S)$ state at a significantly higher
mass \cite{BelleEtacP}, later confirmed by 
the CLEO and BaBar experiments \cite{CLEOEtacP,BaBarEtacP}.  

In this paper we present an investigation of the inclusive photon 
spectrum with the CLEO III detector
from $1.6\cdot 10^6$ $\psi(2S)$ decays,
which is comparable 
in number of resonant decays to the Crystal Ball sample.
The CLEO III detector is equipped with
a CsI(Tl) calorimeter, first 
installed in the CLEO II detector \cite{CLEOII},
with energy resolution matching that of the Crystal Ball detector.
The finer segmentation of the CLEO calorimeter provides
for better photon detection efficiency and more effective
suppression of the photon background from $\pi^0$ decays.
The CLEO III tracking detector, 
consisting of 
a silicon strip detector and a large drift chamber \cite{CLEOIIIDR}, 
provides improved suppression of backgrounds 
from charged particles.
The magnetic field inside the tracking detector was 1 T. 

The data used in this analysis were collected at the
CESR $e^+e^-$ storage ring, which
operated for over two decades at the $b\bar b$ threshold energy 
region. Recently, CESR has been reconfigured to 
run near the $c\bar c$ threshold by insertion of 
12 superconducting wiggler magnets. The data analyzed in
this article come from the first stage of this upgrade
in which the first superconducting wiggler magnet was 
installed. The peak instantaneous luminosity 
achieved was $2\cdot10^{31}$ cm$^{-2}$s$^{-1}$.
The integrated luminosity collected and analyzed
at the $\psi(2S)$ peak region  was 2.7 pb$^{-1}$.

\def\ntk{N_{ch}}
The data analysis starts with the selection of 
hadronic events detected at the $\psi(2S)$
resonance. We require that the observed number of charged tracks ($\ntk$) 
be at least one.
The visible energy of tracks and photons 
($E_{vis}$) must be 
at least 20\%\ (40\%\ if $\ntk=1$)
of the center-of-mass energy ($E_{CM}$).
For $1\le\ntk\le3$ the total energy visible in the 
calorimeter alone ($E_{cal}$) must be at least 15\% of 
$E_{CM}$. 
To suppress $\ee\to\ee$ and $\ee\to\mm$ events, the 
momentum of the second most energetic track in the
event must be less than 85\%\ of the beam energy,
and $E_{cal}<0.85\,E_{CM}$ for $\ntk\le3$.
We also veto events with invariant mass of 
$e^+e^-$ or $\mu^+\mu^-$ within 100 MeV of the
$J/\psi(1S)$ mass to avoid statistical correlations between 
this analysis and our studies of the two-photon cascades 
in $\gamma\gamma l^+l^-$ events.
The lepton pair candidates are selected using energy
deposited in the calorimeter.
The resulting event selection 
efficiency is $84\%$ for decays of 
the $\psi(2S)$ resonance.

\begin{figure}[htbp]
\vbox{
\includegraphics*[width=3.5in]{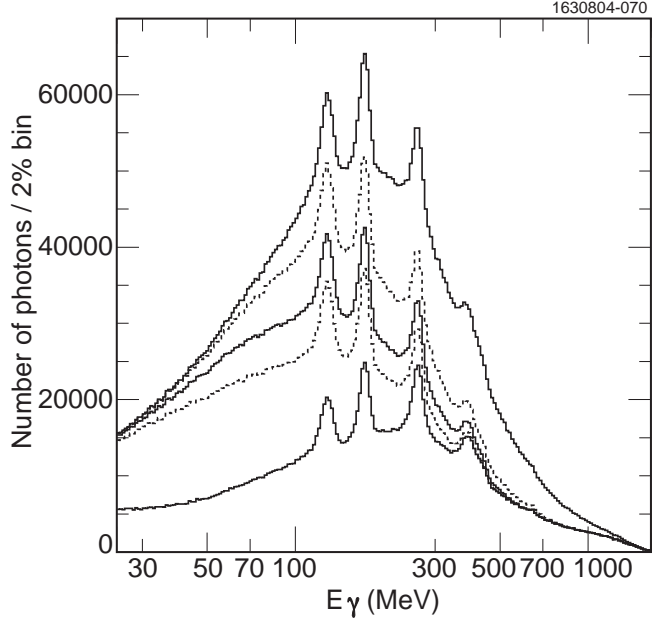}
\caption{
Going from bottom to top, the histograms show the photon energy
spectra at the $\psi(2S)$ resonance 
omitting those photons which, with another photon in the event,
form a $\pi^0$ candidate with  $\cos\theta_{\gamma\gamma}$
exceeding:
-1; i.e.\  maximal $\pi^0$ suppression (solid),
0.3 (dashed), 0.5 (solid), 0.7 (dashed),
1.0; i.e.\ no $\pi^0$ suppression (solid).
The first three peaks correspond to E1 photon transitions:
$\psi(2S)\to\gamma\chi_{cJ}(1P)$, $J=2$ ($128$ MeV), 
$1$ ($172$ MeV) and $0$ ($262$ MeV).
The fourth peak around $400$ MeV 
is an overlap of two E1 photon lines: 
$\chi_{cJ}(1P)\to\gamma J/\psi(1S)$, $J=1$ and $2$.
A small peak at $646$ MeV is due to hindered M1
photon transition: $\psi(2S)\to\gamma\eta_c(1S)$.
}
}
\label{fig:pizvar}
\end{figure}

In the next step of the data analysis we select photon candidates. 
Showers in the calorimeter are required not to match the
projected trajectory of any charged particle, and
to have lateral shower profile consistent with
that of an isolated electromagnetic shower.
We restrict the photon candidates to 
be within the central barrel part of the calorimeter
($|\cos\theta|<0.8$) where the photon energy resolution is
optimal. 
The main photon background in this analysis comes from
$\pi^0$ decays. We can reduce this background by removing 
photon candidates that combine with another photon 
to fit the $\pi^0$ mass within the experimental resolution. 
Unfortunately, this lowers the signal efficiency, 
since random photon combinations sometime fall within the 
$\pi^0$ mass window. The number of random matches to the 
$\pi^0$ hypothesis can be decreased by restricting the 
opening angle between the two photons ($\theta_{\gamma\gamma}$).
Fig.~\ref{fig:pizvar} shows photon energy spectra obtained with various 
levels of
$\pi^0$ suppression. 
We choose the cut $\cos\theta_{\gamma\gamma}>0.5$ to optimize the 
statistical sensitivity in the
widest range of photon energies. 

\begin{figure}[htbp]
\includegraphics*[width=3.5in]{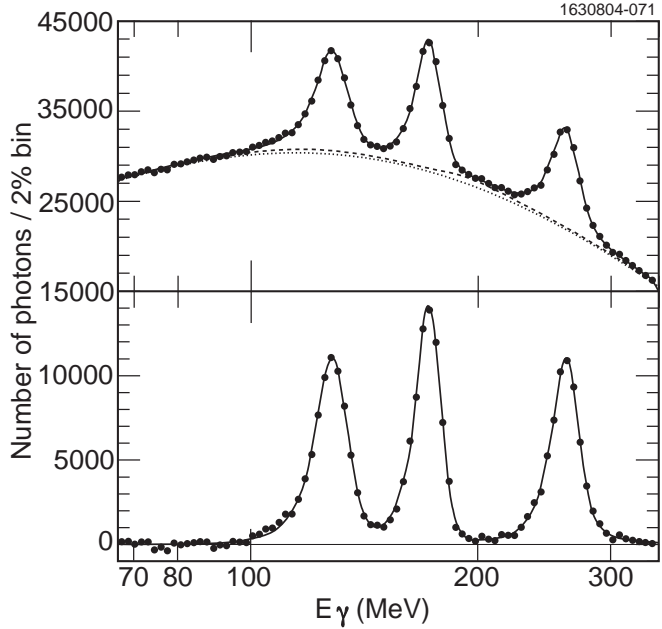}
\caption{
Fit of $\psi(2S)\to\gamma\chi_{cJ}(1P)$ ($J=2,1,0$) photon
lines to the data.
The points represent the data.
Statistical errors on the data 
are smaller than the point size.
The solid line represents the fit.
The dashed line represents total fitted
background. 
The dotted line 
represents the polynomial background alone,
without the contributions from charged particles and
the decays $J/\psi(1S)\to\gamma\eta_c(1S)$.
The background subtracted data (points) 
and the fitted photon lines superimposed 
(solid line) are shown at the bottom. 
}
\label{fig:lnenominalfit}
\end{figure}

Fig.~\ref{fig:lnenominalfit} shows the fit to the three dominant E1 photon 
lines $\psi(2S)\to\gamma\chi_{cJ}(1P)$ ($J=2,1,0$). 
Each photon line is represented 
by a convolution of a non-relativistic
Breit-Wigner  and a detector response
function. The latter is parametrized by the 
so-called Crystal Ball line shape, which 
is a Gaussian (described by the peak energy, 
$E_p$, and energy resolution, $\sigma_E$) 
turning into a power law tail, $1/(E_p-E+const)^n$,
at an energy of $E_p-\alpha\,\sigma_E$.
This asymmetric low-energy tail is induced by
the transverse and longitudinal shower energy
leakage out of the group of crystals used in the
photon energy algorithm.
We determine the parameters describing the leakage tail, 
$n$ and $\alpha$, from the fit 
to $\psi(2S)\to\gamma\chi_{cJ}(1P)$ photon lines observed
with essentially no background
in $\gamma\gamma l^+l^-$ events (produced via
subsequent
$\chi_{cJ}(1P)\to\gamma J/\psi(1S)$, $J/\psi(1S)\to l^+l^-$
decays).
The natural widths of the $\chi_{cJ}(1P)$ states 
are fixed to their world average values \cite{PDG}.
Energy resolution parameters, $\sigma_E$,
are allowed to float. 
The energy resolution dominates over the natural
widths, although the contribution of the natural width 
is significant for the $J=0$ line.
Averaging over the three fitted peaks and using the 
photon energy dependence predicted by the Monte
Carlo, we get a fitted photon energy resolution 
(extrapolated to $E_\gamma=100$ MeV) of 
$4.8\pm0.3$ MeV, where the error is systematic. 
We represent the photon background under the peaks 
by a $4^{th}$ order polynomial.
We include also the charged particle energy 
distribution measured for photon candidates
matched to charged tracks, with the track-match-miss  
probability fixed to the expected
value (1\%). The peaking of this distribution 
around 200 MeV comes from minimum 
ionizing tracks. 
We also include in the fit a small Doppler broadened photon line 
at 115 MeV due to $\psi(2S)\to X J/\psi(1S)$,
$J/\psi(1S)\to\gamma\eta_c(1S)$ with the
amplitude fixed to the number estimated 
using the world average branching ratios \cite{PDG}.

The fitted number of events, with photon line energies 
and statistical errors are 
$79\,300\pm1\,180$ ($128.00\pm0.08$ MeV),
$76\,700\pm  910$ ($172.05\pm0.08$ MeV) and
$72\,630\pm  930$ ($261.99\pm0.14$ MeV) for
$\psi(2S)\to\gamma\chi_{cJ}(1P)$ $J=2$, $1$ and $0$,
respectively. 

To estimate systematic errors on the fitted photon 
energies we vary the fitted range, the order of 
the background polynomial, the detector response
parameters ($\alpha$ and $n$), the normalization of the 
charged particle spectrum, 
the natural widths of $\chi_{cJ}(1P)$ states; 
and we use linear rather than logarithmic  
binning in energy. We also vary the level of 
$\pi^0$ suppression from no suppression 
to $\cos\theta_{\gamma\gamma}>0.3$
(see Fig.~\ref{fig:pizvar}).
The corresponding uncorrelated 
systematic errors on the photon energies
are $\pm0.08\%$, $\pm0.10\%$ and $\pm0.13\%$ 
for $J=2$, $1$ and $0$, respectively.
There is also a common energy scale error of
$\pm0.5\%$, due to the systematic limitation 
of the calibration procedure based on 
$\pi^0$ and $\eta$ masses \cite{calibration}.
Adding the statistical and uncorrelated 
systematic errors together (the first error) and
factoring out the energy scale error (the second
error), we get the following measured photon energies: 
$(128.00\pm0.13\pm0.64)$ MeV,
$(172.05\pm0.19\pm0.86)$ MeV, and 
$(261.99\pm0.37\pm1.31)$ MeV 
for $J=2$, $1$ and $0$, respectively, 
in good agreement with the masses of 
the $\psi(2S)$ and $\chi_{cJ}(1P)$ states 
precisely measured via 
scans of these resonances \cite{PDG}.
A ratio of the fine mass splitting in the
$\chi_{cJ}(1P)$ triplet determined from our
photon energy measurements is:
$r\equiv (M(\chi_{c2})-M(\chi_{c1}))/(M(\chi_{c1})-M(\chi_{c0}))$
$=0.490\pm0.002\pm0.003$.

The fitted peak amplitudes serve for the determination of 
the photon transition branching ratios.
We determine signal selection efficiencies by 
Monte Carlo simulation to be 
$54\%, 54\%$ and  $50\%$ for $J=2, 1$ and $0$, including
factors due to spin-dependent $\cos\theta$ distributions.
To obtain branching ratios, we divide the
efficiency-corrected photon yields 
by the number of $\psi(2S)$ resonances produced.
The latter is determined by the background-subtracted and
efficiency-corrected yield of hadronic events in our
data. To estimate systematic uncertainty, 
hadronic event selection criteria are varied, resulting in a
$\psi(2S)$ efficiency change from $58\%$ to 
$87\%$. Cosmic ray, beam-gas and beam-wall backgrounds
vary from $0.1\%$ to $3.3\%$ as
estimated from the tails of the 
event vertex distribution along
the beam direction. The QED backgrounds
($e^+e^-\to e^+e^-$, 
$e^+e^-\to \mu^+\mu^-$ and 
$e^+e^-\to \tau^+\tau^-$)
are estimated using Monte Carlo
simulations normalized to theoretically
calculated cross-sections \cite{QED} 
and measured integrated luminosity.
The Bhabha scattering
background varies from 0 to $2\%$, while the
$\mu-$pair and $\tau-$pair backgrounds
do not exceed $0.3\%$. 
Continuum production of hadrons
amounts to a $2.4-2.6\%$ background
subtraction and is estimated 
using the Monte Carlo simulation 
normalized to the continuum cross-section measured
by the other experiments \cite{PDG}.
Uncertainties in modeling of
decays of $c\bar c$ states partially cancel
between the signal and hadronic selection
efficiencies. The overall systematic error
in the branching ratio normalization is $\pm3\%$.   
The other systematic errors on the branching ratios are 
determined by the photon selection and fit variations described previously. 
Electromagnetic shower simulations contribute an additional
$\pm2\%$.

Our ${\cal B}(\psi(2S)\to\gamma\chi_{cJ}(1P))$ results are 
 $(9.33\pm0.14\pm0.61)\%$,
 $(9.07\pm0.11\pm0.54)\%$, and 
 $(9.22\pm0.11\pm0.46)\%$ 
for $J=2$, $1$ and $0$, respectively.
They are significantly higher than values
obtained by the Particle Data Group by
a global fit to the $\psi(2S)$ data \cite{PDG}, 
but agree well with the previous
measurements by the Crystal Ball \cite{CB86}, 
($8.0\pm0.5\pm0.7$,
 $9.0\pm0.5\pm0.7$ and 
 $9.9\pm0.5\pm0.8$, in percent, respectively),
and have improved statistical and 
systematic errors.
Since the statistical and systematic errors are
partially correlated for the three
$\chi_{cJ}(1P)$ states we also provide
a sum and ratios of these branching ratios
with properly evaluated errors:
${\cal B}(\psi(2S)\to\gamma\chi_{c0,1,2})(1P)=(27.6\pm0.3\pm2.0)\%$,
${\cal B}(\psi(2S)\to\gamma\chi_{c2}(1P))/
 {\cal B}(\psi(2S)\to\gamma\chi_{c1}(1P))=1.03\pm0.02\pm0.03$,
${\cal B}(\psi(2S)\to\gamma\chi_{c0}(1P))/
 {\cal B}(\psi(2S)\to\gamma\chi_{c1}(1P))=1.02\pm0.01\pm0.07$ and
${\cal B}(\psi(2S)\to\gamma\chi_{c0}(1P))/
 {\cal B}(\psi(2S)\to\gamma\chi_{c2}(1P))=0.99\pm0.02\pm0.08$.

\begin{figure}[htbp]
\includegraphics*[width=3.5in]{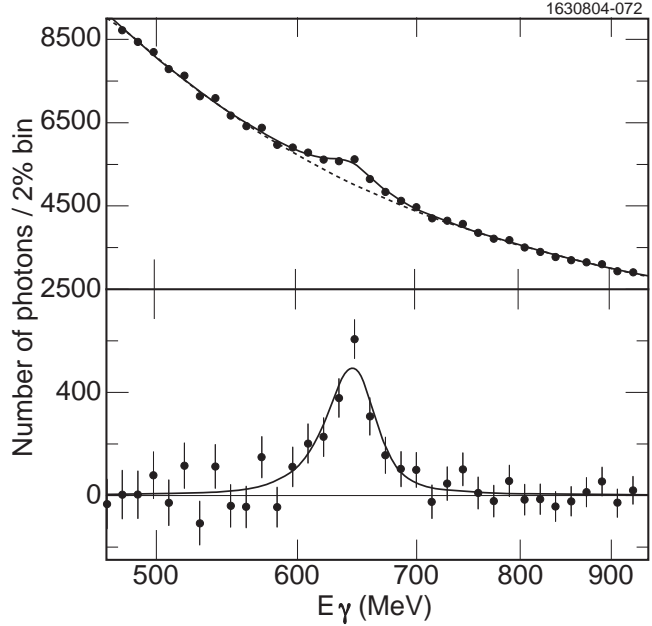}
\caption{
Fit of $\psi(2S)\to\gamma\eta_c(1S)$ photon line to the
data. 
The points represent the data.
The solid line represents the fit.
The dashed line represents the total fitted
background. 
The background subtracted data (points with error
bars) and the fitted photon line superimposed 
(solid line) are shown at the bottom.  
}
\label{fig:etac}
\end{figure}

In addition to the dominant E1 photon peaks, we 
also observe a small peak at higher photon energy 
due to the hindered M1 transition $\psi(2S)\to\gamma\eta_c(1S)$. 
The fit, illustrated in Fig.~\ref{fig:etac},
of a Breit-Wigner convoluted with the 
Crystal Ball line shape yields 
$2\,560\pm315$ events
(with a statistical significance of $8.1$ standard deviations) 
and a transition photon energy of $(646.2\pm2.6\pm4.8)$ MeV, where 
the first error is statistical and the second error is
systematic.
The measured photon energy is within $(1.2\pm0.9)\%$ of the
expected value determined from the world average
masses of the $\psi(2S)$ and 
$\eta_c(1S)$ states \cite{lp03-skwa}.

The detection efficiency is $51\%$ for this transition. 
The fitted peak amplitude depends strongly on the assumed
natural width of the $\eta_c(1S)$. 
In our nominal fit we assumed $\Gamma_{\eta_c(1S)}=24.8\pm 4.9$ MeV, 
coming from our own determination 
via formation in $\gamma\gamma$ fusion \cite{CLEOEtacP}.
When left free in the fit, the fitted width 
is consistent with this value.
Since the exact value of this width is a subject of 
experimental controversy \cite{lp03-skwa}, 
we factor the $\Gamma_{\eta_c(1S)}$ dependence out to
enable a rescaling of our results to a different value in 
the future.
The central value of the branching ratio
(${\cal B}$) can be expressed
as 
$$
{\cal B}
=
\left(0.324+0.028\,\frac{\Gamma_{\eta_c(1S)}-24.8{\rm  MeV}}{4.9{\rm  MeV}}
\right)\quad\%
$$
and the errors are
$$
\left( 
\pm0.039\pm0.055
\right) \frac{{\cal B}
}{0.324\%}\quad
\pm\left(0.028\,\frac{\Delta\Gamma_{\eta_c(1S)}}{4.9{\rm  MeV}}\right)\quad
\%.
$$
The first error is statistical, the second is systematic, and the 
third is due to the uncertainty in the $\Gamma_{\eta_c(1S)}$ width.
For our nominal choice of $\Gamma_{\eta_c(1S)}$ and
$\Delta\Gamma_{\eta_c(1S)}$ we obtain 
${\cal B}(\psi(2S)\to\gamma\eta_c(1S))=(0.32\pm0.04\pm0.06)\%$.
The first error is statistical and the second is the total 
systematic uncertainty.
This is the first confirmation of this transition, previously 
detected by the Crystal Ball.
The Crystal Ball measured the rate for this transition
to be $(0.28\pm0.06)\%$ for $\Gamma_{\eta_c(1S)}=(11.5\pm4.5)$ MeV \cite{CB86}.
Rescaled to this width, our result, $(0.25\pm0.06)\%$, is 
in good agreement with their measurement.

The Crystal Ball also presented evidence for the direct  \cite{CB82}
M1 transition $\psi(2S)\to\gamma\eta_c(2S)$ with
$E_\gamma=(91\pm5)$ MeV, 
${\cal B}(\psi(2S)\to\gamma\eta_c(2S))$ 
in the 95\%\ confidence level (C.L.) interval
$(0.2-1.3)\%$
and $\Gamma_{\eta_c(2S)}<8$ MeV (95\%\ C.L.).
We see no evidence for such a photon line
(see Fig.~\ref{fig:lnenominalfit})
and set an upper limit for the transition
rate to a $\Gamma=8$ MeV state at this
photon energy to be $<0.2\%$ at 90\%\ C.L. 

Recent observations of the $\eta_c(2S)$ state \cite{lp03-skwa}
imply the photon energy for such a transition 
to be about $47$ MeV with a larger 
width. Since a small and 
wide photon line at such low
photon energy 
cannot be distinguished
from the photon backgrounds
(for example, the width of such a peak for $\Gamma_{\eta_c(2S)}=25$ MeV 
would be about 4 times broader
in Fig.~\ref{fig:pizvar} than the width of
the first E1 line),
we have no meaningful sensitivity for such a 
transition at the new $\eta_c(2S)$ mass.

In summary, we have improved branching ratio measurements 
for $\psi(2S)\to\gamma\chi_{c0,1,2}(1P)$ (E1 transitions)
and $\psi(2S)\to\gamma\eta_c(1S)$ (hindered M1 transitions).
The latter is the first confirmation of the existence of such a 
transition. The direct M1 transition $\psi(2S)\to\gamma\eta_c(2S)$
is not observed and the upper limit is set below the 
branching ratio range previously claimed by the
Crystal Ball experiment \cite{CB82}. 

We gratefully acknowledge the effort of the CESR staff in providing us with
excellent luminosity and running conditions.
This work was supported by 
the National Science Foundation and
the U.S. Department of Energy.


\end{document}